\documentclass[amssymb,aps,prl,twocolumn,amsmath,groupedaddress]{revtex4}
\usepackage{graphicx}

\begin{document}

\title{Coherent control of optical four-wave mixing by 
two-color $\omega$-$3\omega$ ultrashort laser pulses}
\author{Carles Serrat}
\affiliation{Departament de F\'{\i}sica i Enginyeria Nuclear, Universitat Polit\`{e}cnica de Catalunya,
Colom 1, 08222 Terrassa, Spain}

\date{\today}

\begin{abstract}

A theoretical investigation on the quantum control of optical coherent 
four-wave mixing interactions in two-level systems driven by two intense 
synchronized femtosecond laser pulses of 
central angular frequencies $\omega$ and $3\omega$ is reported.    
By numerically solving the full Maxwell-Bloch equations beyond the slowly-varying 
envelope and rotating-wave approximations in the time domain, 
the nonlinear coupling to the optical field at frequency $5\omega$ is
found to depend critically on the initial relative phase $\phi$ 
of the two propagating pulses; 
the coupling is enhanced when the pulses interfere constructively in the center
($\phi=0$), while it is nearly suppressed when they are out of phase ($\phi=\pi$). 
The tuning of the initial absolute phase of the different 
frequency components of synchronously propapagating $\omega$-$3\omega$ femtosecond 
pulses can serve as a means to control coherent anti-Stokes Raman (CARS) processes. 

\end{abstract}


\maketitle

In recent years, encouraged by the developments in the engineering of intense ultrashort laser fields 
with a well defined absolute phase \cite{Paulus}, studies on the phase 
control of the interaction of two-color strong ultrashort laser
pulses in nonlinear materials have received a great interest 
\cite{photodiss,Watanabe,photoion,Bandrauk,Brown,Xu}. 
Phenomena arising from such ultrashort pulse interaction can be of extreme importance 
in fields as diverse as optoelectronics and materials research, in biological
applications such as spectroscopy and microscopy, 
in high harmonic generation, 
and in photoionization 
or molecular dissociation, 
among others.

It is known that when the pulses duration approach the duration of only several optical cycles, 
theories based on the slowly-varying envelope approximation (SVEA) and the rotating-wave approximation 
(RWA) are not appropriate, since phenomena such as electric field time-derivatives leading to 
carrier-wave reshaping \cite{Ziolkowski}, or the generation of high spectral components \cite{Hughes}, 
can not be described by such theories. 
In these situations, accurate numerical modeling based on
the finite-difference time-domain (FDTD) method \cite{Taflove} is being increasingly used 
to investigate the full set of optical Maxwell-Bloch equations \cite{Xu,Ziolkowski,Hughes,Tarasishin}. 

In this Letter, we investigate the possibility of 
phase control of transient four-wave mixing interactions in two-level 
systems driven by two mutually coherent intense ultrashort pulses of central angular 
frequencies $\omega$ and $3\omega$. 
We employ a standard predictor-corrector FDTD numerical approach 
which incorporates all propagation effects --such as nonlinearity, dispersion, 
absorption, dissipation, saturation, and other resonance effects-- \cite{Ziolkowski}, 
to study the sensitivity on the relative phase $\phi$
of the nonlinear coupling of the $\omega$-$3\omega$ pulses to the field
at frequency $5\omega$,
which results from the interaction of the waves through the third order
susceptibility of the medium ($\chi^{(3)}$) \cite{Boyd}.

\begin{figure}[]
\begin{center}
\includegraphics[scale=0.39,clip=true,angle=270]{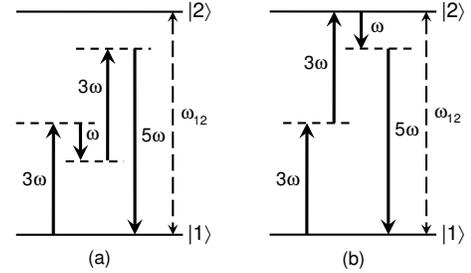}
\caption{\label{fig1} 
Schematic energy level diagrams corresponding to the first case of our study. 
The four-wave mixing can occur between the injected two-color 
fields at frequencies $\omega$ and $3\omega$, 
and a third generated field at frequency $5\omega$, by parametric coupling (a) or
by two-photon processes (b). 
The wavelengths associated to the frequencies in this scheme are:
$\lambda_{(\omega)}=2400$ nm; $\lambda_{(3\omega)}=800$ nm; 
$\lambda_{(5\omega)}=480$ nm; $\lambda_{(\omega_{12})}=400$ nm.
The area of both pulses is $A=20\pi$.}
\end{center}
\end{figure}
By describing the evolution of the field spectrum as the two-color pulses propagate 
through the medium, our simulations demonstrate that the generation of the $5\omega$-component
is enhanced when the two pulses are in phase ($\phi=0$), while it is nearly suppressed when they are out 
of phase ($\phi=\pi$).
In what follows, we analyze this effect in two different configurations: \\
\hspace{3cm} -- First, we consider a nonresonant case with large 
pulse intensities (see Fig. \ref{fig1}). 
The two injected pulses have a duration of $10$ fs and they
initially overlap in time. The pulse area for both pulses is $A=20\pi$ \cite{area}, and the
wavelengths are $\lambda_{(\omega)}=2400$ nm and $\lambda_{(3\omega)}=800$ nm.
The fields are largely detuned from the  
atomic transition resonance ($\lambda_{(\omega_{12})}=2\pi c/\omega_{12}=400$ nm),
although the two-photon coupling of the field with $\lambda_{(3\omega)}=800$ nm is considered in resonance 
(see Fig. \ref{fig1} (b)). Therefore, 
by four-wave mixing, the fields can interact with a generated third wave with
$\lambda_{(5\omega)}=480$ nm, by either the
parametric coupling represented in Fig. \ref{fig1} (a) or through a two-photon process [Fig. \ref{fig1} (b)].
We note that the energy level diagrams of this first study can be considered as an example of the vibrational nonresonant
contributions at zero pump-probe time delay in CARS processes \cite{CARS}. \\
\hspace{3cm} -- In a second study, we consider two $10$ fs pulses with moderate pulse areas ($A=4\pi$) 
and wavelengths as
$\lambda_{(\omega)}=1743$ nm and $\lambda_{(3\omega)}=581$ nm.
In this case, the field with $\lambda_{(3\omega)}=581$ nm propagates close to resonance 
with the atomic transition ($\lambda_{(\omega_{12})}=580$ nm).
By a nonlinear third order parametric coupling (see Fig. \ref{fig3})
the fields can produce a wave with $\lambda_{(5\omega)}=348.6$ nm.
The parameter values in this second study have been chosen to be close 
to those for transitions that are relevant in
biological applications, with similar schemes being currently investigated with 
femtosecond CARS techniques \cite{CARS,fsCARS,Yaron}.

The Maxwell-Bloch equations can be written as \cite{Ziolkowski} 
\begin{eqnarray}\label{MBS}
\frac{\partial{H_y}}{\partial{t}}&=&-\frac{1}{\mu_0}\frac{\partial{E_x}}{\partial{z}} \nonumber \\
\frac{\partial{E_x}}{\partial{t}}&=&-\frac{1}{\epsilon_0}\frac{\partial{H_y}}{\partial{z}}-
\frac{N_{at}\Gamma}{\epsilon_0 T_2}(\rho_1-T_2\omega_{12}\rho_2) \nonumber \\
\frac{\partial{\rho_1}}{\partial{t}}&=&-\frac{1}{T_2}\rho_1 +\omega_{12}\rho_2 \\
\frac{\partial{\rho_2}}{\partial{t}}&=&-\frac{1}{T_2}\rho_2 +\frac{2\Gamma}{\hbar}E_x\rho_3-\omega_{12}\rho_1 \nonumber \\
\frac{\partial{\rho_3}}{\partial{t}}&=&-\frac{1}{T_1}(\rho_3-\rho_{30})-\frac{2\Gamma}{\hbar}E_x\rho_2 \nonumber 
\end{eqnarray}
where $H_y(z,t)$ and $E_x(z,t)$ represent the magnetic and electric fields
propagating along the $z$ direction, respectively,
$\mu_0$ and $\epsilon_0$ are the magnetic permeability and 
electric permittivity of free space,  respectively,
$N_{at}$ represents the density of polarizable atoms,
$\Gamma$ is the dipole coupling coefficient,
$T_1$ is the excited-state lifetime, $T_2$ 
is the dephasing time,
$\omega_{12}$ is the transition resonance angular frequency of 
the two level medium, and 
$\rho_1$ and $\rho_2$ are the real and imaginary
components of the polarization --which determine the index of refraction and gain coefficients, 
respectively-- \cite{Ziolkowski}.
The population difference is $\rho_3$, and $\rho_{30}$ represents its initial value.
\begin{figure}[!h]
\begin{center}
\includegraphics[scale=0.45,clip=true,angle=270]{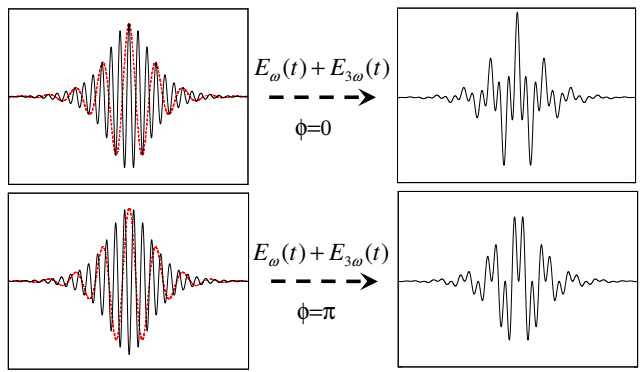}
\caption{\label{fig15} 
Superposition of the electric field of the pulses. 
$E_\omega(t)$ is the electric field of the pulse with frequency 
$\omega$ represented as a function of time, and $E_{3\omega}(t)$ represents the pulse
with frequency $3\omega$ (see also Eq. (\ref{EINJ}). As indicated, the upper figures 
correspond to the addition of the pulses in phase ($\phi=0$), and the
lower ones correspond to the addition out of phase ($\phi=\pi$).}
\end{center}
\end{figure}

As mentioned above, an hyperbolic secant two-color pulse,
which can be expressed as
\begin{eqnarray}
&E_x(z=0,t)=E_\omega(t)+E_{3\omega}(t)= E_0sech((t-t_0)/t_p)\times &  \nonumber \\
&\left[cos(\omega (t-t_0))+cos(3\omega (t-t_0)+\phi)\right],& \label{EINJ}
\end{eqnarray}
is externally injected to the system.
The peak input electric field amplitude $E_0$
is chosen the same for both pulses and, in terms of the pulse area $A$, 
it is given by $E_0=A\hbar/(\pi \Gamma t_p)$ \cite{area}. 
The duration of the pulse is given by $t_p=\tau_p /1.763$, 
with $\tau_p$ being the full width at half maximum (FWHM) of the
pulse intensity envelope. $t_0$ gives the offset position of the
pulse center at $t=0$, and it is the reference value for the 
absolute phase of the pulses. The angular frequencies of the 
pulses are $\omega$ and $3\omega$, and $\phi$ is the relative phase. 
The parameters have been chosen as follows:
$T_1=T_2=1$ ns,
$\tau_p=10$ fs,
$\Gamma=2.65e$\r{A},
 and $N_{at}=2\times10^{18}$ cm$^{-3}$.
It is important to note that our simulations give information
on the coherent (transient) behavior of the system, since 
the dephasing time scales are chosen much larger than the duration of the 
pulses $T_1,T_2>>\tau_p$.

Figure \ref{fig15} is a representation of the electric field of the injected 
pulses. In the two upper figures we can observe the addition of two pulses 
with frequencies $\omega$ and $3\omega$ in the case that the relative phase $\phi$ between
them is $\phi=0$. We note that the constructive interference in the
center of the pulses results in a maximum of the resulting pulse (Fig. \ref{fig15} upper-right), 
while in the case that the pulses are added out of phase ($\phi=\pi$), there is destructive interference
in the center of the pulses and therefore 
the resulting pulse (Fig. \ref{fig15} upper-right) presents a zero in the center.  
We will show below that the different shape, which only depends on the
relative phase between the two $\omega-3\omega$ pulses, is crucial in the 
nonlinear interaction
with the two-level medium in the case of ultrashort intense pulses.  

\begin{figure}[!h]
\begin{center}
\includegraphics[scale=0.68,clip=true]{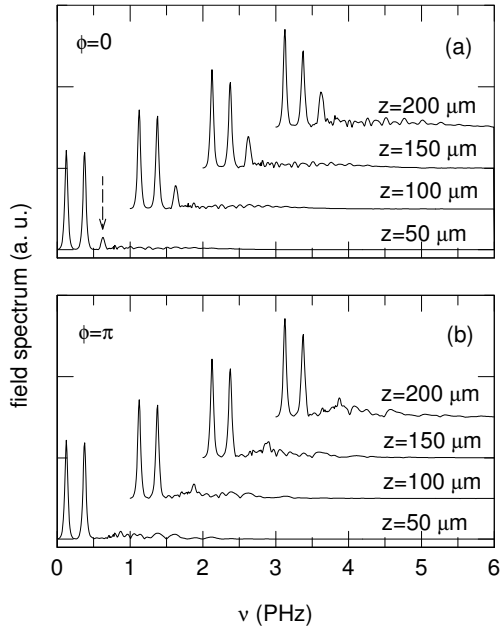}
\caption{\label{fig2} 
Evolution of the field spectra as the pulses propagate through the two-level medium,
as indicated. (a) $\phi=0$; (b) $\phi=\pi$. The dotted arrow in (a) indicates
the peak at frequency $5\omega$.
The spectra have been shifted right and up for different $z$.}
\end{center}
\end{figure}

We next show the results of the simulations corresponding to the first 
scheme (Fig. \ref{fig1}) that we have analyzed. In Fig. \ref{fig2}, 
we represent the spectrum of the
field for different positions of the propagation. The two
highest peaks on the left of each spectra, in Fig. \ref{fig2}, 
correspond to the frequencies of the pulses initially injected to the medium 
($\omega/2\pi$ and $3\omega/2\pi$).
When the two pulses are initially in phase ($\phi=0$), a third frequency (anti-Stokes) component
 is generated at $5\omega=3\omega+(3\omega-\omega)$, as shown in Fig. \ref{fig2}(a). 
It is clear in that case that the conversion of the third harmonic component ($3\omega$) 
to the $5\omega$-component [which is marked with a dotted arrow in Fig. \ref{fig2}(a)] 
increases as the propagation length increases. 
Differently, however, when the two pulses are initially out of phase ($\phi=\pi$),
the conversion to the $5\omega$-component is completely
suppressed, as it can be observed in Fig. \ref{fig2}(b).
This contrasting behavior, to the 
best of our knowledge, has not been observed before. 
We attribute this phenomenon to 
the dependence on the shape of the carriers (see Fig. \ref{fig15}) 
of the quantum interferences of the medium, which 
in the case of intense ultrashort pulses
are governed by the electric field itself rather than by its envelope \cite{Hughes}. 
\begin{figure}[!h]
\begin{center}
\includegraphics[scale=0.46,clip=true,angle=270]{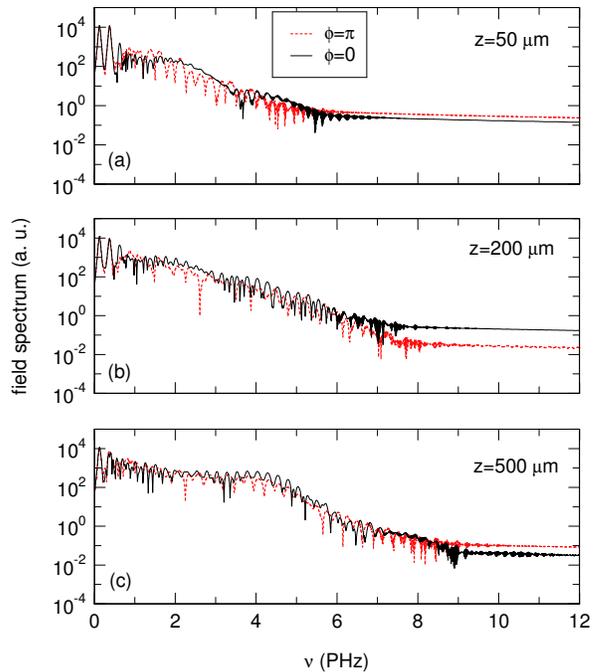}
\caption{\label{fig25} 
Evolution of the field spectra as the pulses propagate through the two-level medium.
(a) $z=50$ $\mu$m; (b) $z=200$ $\mu$m; (c) $z=500$ $\mu$m. 
Note that the verical axis is shown in logarithmic scale.}
\end{center}
\end{figure}

For longer propagation distances ($z \approx 500$ $\mu$m), the
coupling between the fields will not longer be effective,
because of changes in the shape of the pulses, as a consequence of the propagation effects, 
and also because of the progressive change in the spectral energy distribution of the field.
In Fig. \ref{fig25}, we have plotted the spectra of the propagating field, in logarithmic scale,
for propagation distances up to $z=500$ $\mu$m.
We can see how the usual higher spectral components are increasingly generated in the field
as the propagation distance increases.
These higher spectral components are due, on the one hand, to the interference between the pulses and the
associated nonlinear processes \cite{Boyd}, and on the other hand, to
carrier-wave Rabi flopping phenomena which are present due to the 
high energies contained in the injected pulses \cite{Hughes}. 
As can be observed in Fig. \ref{fig25}(c), in the case of the present simulations, the spectrum 
spreads over the entire UV region for $z=500$ $\mu$m. Note that
some sensitivity on the initial relative phase $\phi$ of the
spectra, in the higher frequency region, can also be observed in Fig. \ref{fig25}. Further details
on these long propagation effects, however, are left to be given elsewhere. 
\begin{figure}[!h]
\begin{center}
\includegraphics[scale=0.41,clip=true,angle=270]{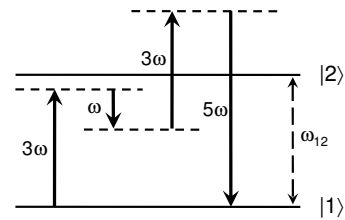}
\caption{\label{fig3}
Energy level scheme representing our second study.
As in the previous scheme (\ref{fig1}), the coupling can occur between the  
fields injected at frequencies $\omega$ and $3\omega$, and a third field generated at $5\omega$.
The wavelengths associated to the frequencies in this scheme are:
$\lambda_{(\omega)}=1743$ nm; $\lambda_{(3\omega)}=581$ nm; 
$\lambda_{(5\omega)}=348.6$ nm; $\lambda_{(\omega_{12})}=580$ nm.
The area of the pulses is $A=4\pi$.
Note that, in this case, the field at $3\omega$
 is close to resonance with the transition.}
\end{center}
\end{figure}

\begin{figure}[!h]
\begin{center}
\includegraphics[scale=0.63,clip=true]{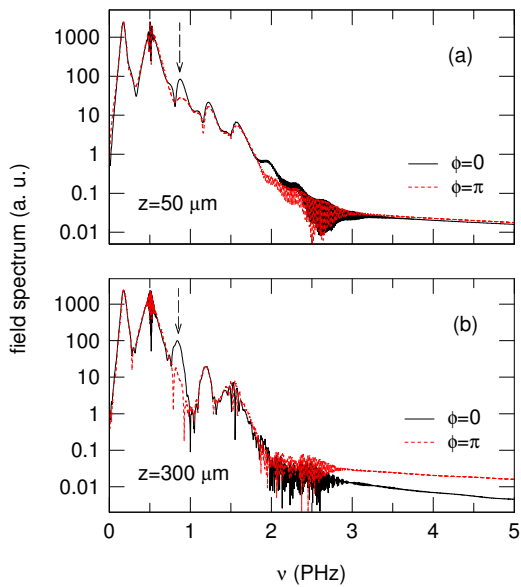}
\caption{\label{fig4} Field spectra at $z=50$ $\mu$m (a) and $z=300$ $\mu$m (b). 
The dotted arrows indicate the peaks at frequency $5\omega$ for $\phi=0$.}
\end{center}
\end{figure}

The second energy level configuration that we have analyzed is represented in
 Fig. \ref{fig3}. In this case, two overlapping pulses
with a duration of $\tau_p=10$ fs are injected to the two-level medium. 
The area of the pulses
is in that case much smaller ($A=4\pi$) than in the previous scheme, and the field at frequency
$3\omega$ is now nearly in resonance with the transition of the medium. 
In Fig. \ref{fig4} the field spectra for two different positions 
are shown ($z=50$ $\mu$m [Fig. \ref{fig4}(a)] and $z=300$ $\mu$m [Fig. \ref{fig4}(b)]). 
We clearly observe that the transfer of energy to the
$5\omega$-component (which is marked with a dotted arrow in Fig. \ref{fig4}) 
is only efficient in the case
of constructive interference in the centre of the pulses ($\phi=0$), while it is almost
suppressed for ($\phi=\pi$). On the other hand, 
the rest of the main peaks in the spectra,
which result from the interference of the fields, remain basically insensitive to the variations 
of $\phi$, and we therefore conclude that the influence of the initial relative phase is 
mostly observed in the four-wave coupling to the $5\omega$-component. 
This second study hence provides another demonstration of the possibilities of the 
phase control phenomena that we report.

We finally note that the results of the second scheme that we have investigated 
can be compared with previous work \cite{Xu}, where
in a similar configuration the production of higher spectral components even for small pulse areas 
was demonstrated, and it was shown that the oscillatory structures around the resonant frequency 
and the propagation features of the laser pulses depend sensitively on the relative phase 
$\phi$ of the two pulses. We have here addressed the influence of the relative phase on the nonlinear 
coupling to the optical field at frequency $5\omega$, which was not considered in \cite{Xu}, and in 
that sense the second study in our work can be considered as complementary to that investigation.

In conclusion, we have analyzed the coherent propagation of two-color $\omega$-$3\omega$ 
femtosecond laser pulses overlapping in time and propagating in a two-level system
to an extend of some hundreds of microns. 
Our study predicts a critical dependence on the relative initial phase for
the transfer of energy by four-wave mixing to the field at frequency $5\omega$. 
This effect is clearly observed in our simulations for propagation
distances as long as $z\approx 300$ $\mu$m. 
We have observed this phenomenon in the case of 
highly intense femtosecond pulses strongly detuned from the transition of the medium, and 
also for pulses with moderate intensity close to resonance with the material transition. 
We have hence demonstrated that 
the manipulation of the initial absolute phase of the different frequency components of 
sinchronously propagating $\omega$-$3\omega$ femtosecond pulses can serve as a means to control 
the nonlinear coupling to the optical field at frequency $5\omega$
 in four-wave mixing interactions. This coherent control can be useful, in particular,
to limit the background nonresonant contributions in coherent anti-Stokes Raman processes. 

Support from the {\it Programa Ram\'on y Cajal} of the
Spanish Ministry of Science and Technology, from
projects BFM2002-04369-C04-03 and FIS2004-02587, 
and from the Generalitat de Catalunya (project 2001SGR 00223), is acknowledged.

{}
\end{document}